# A Novel RIS-Aided EMF-Aware Beamforming Using Directional Spreading, Truncation and Boosting

Nour Awarkeh*, Dinh-Thuy Phan-Huy†, Raphael Visoz†, Marco Di Renzo*‡

*CentraleSupélec, France, †Orange Innovation, Châtillon, France, ‡CNRS, Université Paris-Saclay, France

*Abstract*— This paper addresses a drawback of massive multiple-input multiple-output Maximum Ratio Transmission beamforming. In some propagation conditions, when the base station serves the same target user equipment for a long period, it reduces the transmit power (and degrades the received power) to avoid creating high exposure regions located in the vicinity of the antenna and concentrated in few directions (corresponding to the best propagation paths between the antenna and the receiver). In this paper, we propose a novel electromagnetic field aware beamforming scheme, which (i) spreads the beamforming radiation pattern in the angular domain by adding artificial propagation paths thanks to reconfigurable intelligent surfaces, (ii) truncates the pattern in strong directions, and (iii) boosts it in weak directions. Compared to existing solutions, it maximizes the received power. However, it also consumes more power. Finally, truncation alone is the best trade-off between received power and energy efficiency, under exposure constrain.

## I. INTRODUCTION

On the one hand, Massive Multiple-Input Multiple-Output (MMIMO) systems and adaptive beamforming (BF) are among the key technologies of mobile networks enabling them to deliver high quality-of-service (QoS) [1][2]. For instance, a Base Station (BS) transmitting with its maximum power $\chi^{max}$ maximizes the received power and the delivered data rate at the target UE thanks to *Maximum Ratio Transmission (MRT)* BF scheme [3] and an MMIMO antenna [4]. On the other hand, the regulation defines a maximum electromagnetic field (EMF) exposure (EMFE) threshold which, must not be exceeded, beyond a predefined region, for instance, a limit circle (in environment without obstacles close to the BS), on average. However, when the BS serves the same user for a long period, in some cases, MMIMO and MRT BF cannot be used or deployed as such, as they could generate an over-exposed area exceeding the limit circle, in some directions [5]-[10]. As illustrated in Fig.1-a), these directions correspond to prime propagation paths between the antenna and the receiver. Also, we foresee that even arbitrarily larger limit circles and more stringent thresholds could be requested in the future by some cities.

One first simple solution to comply with the regulation consists in using a reduced transmit power $\chi^{red} < \chi^{max}$ at the BS (whilst keeping using MRT BF) that ensures that the entire over-exposed area remains inside the circle, even in its strongest directions, and for a long period [11]. Unfortunately, as illustrated in Fig. 1-b), such a *Reduced MRT* BF scheme reduces the received power at the target UE and degrades the received QoS.

To overcome this drawback, a new EMF aware BF scheme, named *Truncated MRT* BF scheme has been recently proposed [11], that truncates the MRT BF radiation pattern, only in the directions where the over-exposed area would exceed the limit circle otherwise. The directions already inside the circle are not impacted by the truncation. Compared to the Reduced MRT BF scheme, the Truncated MRT BF scheme uses a transmit power $\chi^{trunc} > \chi^{red}$ that is higher and delivers a received power at the target UE that is stronger, whilst remaining compliant with the EMFE constrain.

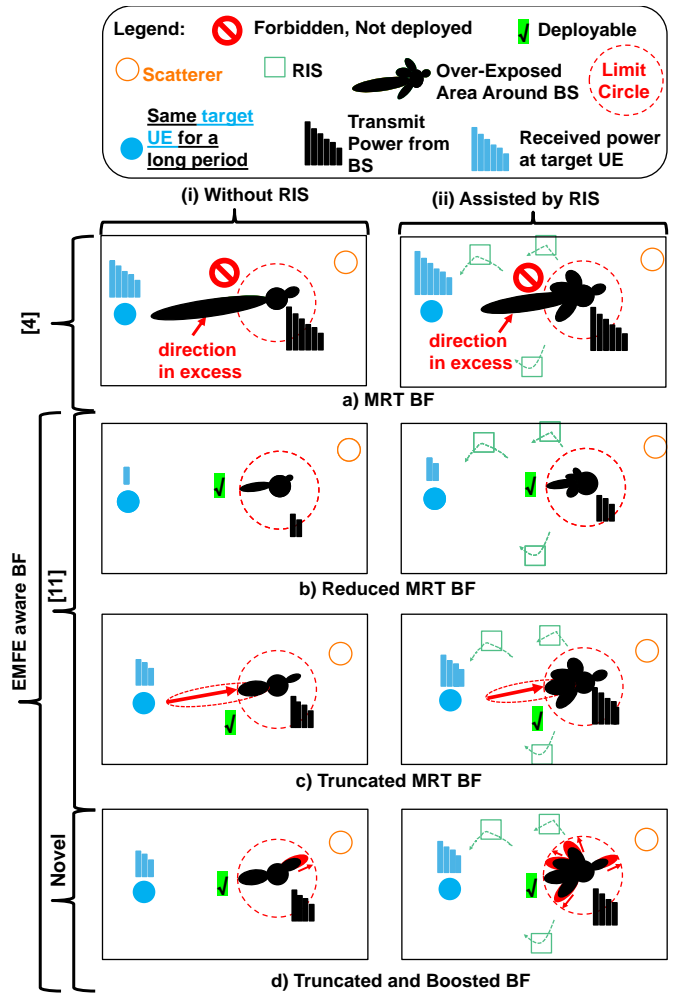

Fig. 1. Visualisation of studied BFs and corresponding over-exposed areas, assuming that the same UE is served by the BS for a long period.

To further improve the performance of the aforementioned BF schemes, [11] has also proposed to exploit the nascent concept of smart radio environments for the future 6[th] generation network (6G), by shaping the propagation environment itself (according to some Channel State Information (CSI)) thanks to reconfigurable intelligent surfaces (RISs) [11]-[16]. Note that in [15][16], RISs are used to reduce self-EMFE reduction due to the smartphone uplink, whereas [11] uses RISs for EMFE reduction in the downlink. More precisely, in [11], a RIS with continuous (instead of discrete) phase-shifting capability as in [17][18] and sensing capability, first measures the propagation channel between the target UE and itself, and then, self-configures to 'turn itself electronically' in the direction of the target UE. In [11], several sensing and self-tuning RISs of such type are deployed in the environment. Hence, without any communication with the RISs, MRT BF, Reduced MRT BF, and Truncated MRT BF schemes (all derived from MRT) naturally spread their radiation patterns in additional directions. As illustrated in Fig. 1-a), 1-b) and 1-c), these directions are the directions of the RISs.

In the current paper, we propose to further improve the schemes proposed in [11], by boosting the BF radiation pattern in the directions where the over-exposed area remains strictly inside the circle. The expected advantage of this novel RIS-aided *Truncated and Boosted MRT BF* scheme is that the transmit power $\chi^{boost} > \chi^{trunc}$ and the received power at the target UE are further improved, as illustrated in Fig. 1-d).

The paper is organized as follows. In Section II, we describe the system model. In Section III, we visualize the impact of the BF schemes on the over-exposed area for a given propagation channel sample. In Section IV, we compare the performances of all schemes based on statistics over randomly generated propagation channels. Finally, we give some concluding remarks in Section V. The following notations are used throughout the paper: $j^2 = -1$, if $x \in \mathbb{C}$, then $\arg(x)$ and $|x|$, are the argument and the module, of $x$, respectively. If $\mathbf{A} \in \mathbb{C}^{M \times N}$, $\mathbf{A}_{m,n}$ is the coefficient of line $m$ and column $n$, with $1 \leq m \leq M$ and $1 \leq n \leq N$, $\|\mathbf{A}\| = \sqrt{\sum_{m=1}^{M}\sum_{n=1}^{N}|\mathbf{A}_{m,n}|^2}$, $\mathbf{A}^\dagger$ and $\mathbf{A}^T$ are the Hermitian and transpose of $\mathbf{A}$. If $\vec{a}, \vec{b} \in \mathbb{R}^{3 \times 1}$ are two points vectors in space with cartesian coordinates, then $\vec{a} \cdot \vec{b}$ is their scalar product. $\mathrm{E}[.]$ is the expectation operation. % denotes percent.

## II. SYSTEM MODEL

We consider the downlink communication between a BS and a target UE. The BS is equipped with an MMIMO antenna consisting of a uniform linear array of $M$ antenna elements spaced by $0.5\lambda$, where $\lambda$ is the wavelength. The target UE has a single antenna. We assume that $K \geq 0$ RIS(s) are deployed in the environment. Note that $K = 0$ corresponds to a scenario without RIS assistance. We assume that each RIS is a uniform linear array of $P$ elements spaced by $0.5\lambda$. Finally, $T \in \mathbb{R}^{3 \times 1}$ denotes the target UE location, $Q$ denotes a position close to the BS, $\omega^{thresh}$ denotes the threshold of received power that must not be exceeded to remain compliant with the EMFE constrain, $\mathcal{C}$ denotes the limit circle around the BS for EMFE, $R$ denotes the radius of $\mathcal{C}$. The BS is assumed to serve the same target UE for a long period.

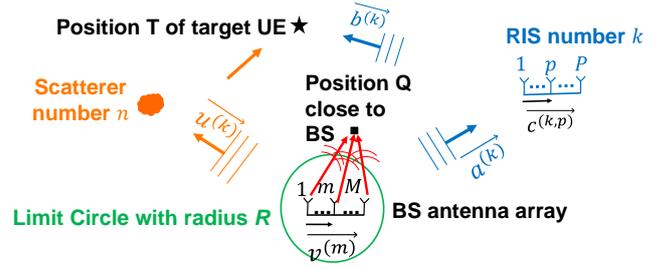

Fig. 2. Propagation channel model

### A. Propagation Channel Model

As illustrated in Fig. 2, we consider a multipath propagation environment between the BS and the target UE, with $N > 1$ scatterers. We assume that an Orthogonal Frequency Division Multiplexing (OFDM) waveform is used and restrict our analysis to a single sub-carrier of the OFDM waveform, as our analysis can be easily generalized to any other sub-carrier of the multi-carrier waveform. With this latter assumption, the propagation channels can be modeled with complex matrices or vectors and the following channel matrices are defined: $\mathbf{s}, \mathbf{h}, \mathbf{g} \in \mathbb{C}^{1 \times M}$ model the multipath propagation channels between the BS and the target UE, through the scatterers, through the RIS(s), through the scatterers and RIS(s)) together, respectively. $\mathbf{q}(Q) \in \mathbb{C}^{1 \times M}$ models the propagation channel between the BS and a UE $Q$ close to (and in line-of-sight of) the BS. With these notations, $\mathbf{g} \in \mathbb{C}^{1 \times M}$ is given by:

$$\mathbf{g} = \mathbf{s} + \mathbf{h}. \tag{1}$$

Note that in the absence of RIS, $\mathbf{h}$ is the null vector and $\mathbf{g} = \mathbf{s}$. The expressions of $\mathbf{s}, \mathbf{h}$ and $\mathbf{q}(Q)$ are provided hereafter.

The scatterers are assumed to be located far away from the BS. Hence, the planar wave approximation applies. With this assumption, $\mathbf{s} \in \mathbb{C}^{1 \times M}$ is given by:

$$\mathbf{s}_m = \sum_{n=1}^{N} \alpha^{(n)} e^{j\frac{2\pi}{\lambda}\left(\overrightarrow{u^{(n)}} \cdot \overrightarrow{v^{(m)}}\right)}, 1 \leq m \leq M, \tag{2}$$

where, $\alpha_n$ is the gain (modeled as a complex random gaussian variable under the Rayleigh fading assumption with $\mathrm{E}[|\alpha_n|^2] = 1$) of the path passing by the $n^{th}$ scatterer, $\overrightarrow{u^{(n)}}$ is the unitary vector indicating the BS-to-$n^{th}$ scatterer direction, and $\overrightarrow{v^{(m)}}$ is the vector between the positions of the 1st and the $m^{th}$ element of the BS. When deployed in the environment, RISs as well, are assumed to be far from the BS and far from the target UE. Hence, again, the planar wave approximation applies. With these assumptions, $\mathbf{h} \in \mathbb{C}^{1 \times M}$ is given by:

$$\mathbf{h}_m = \sum_{k=1}^{K} \frac{\beta^{(k)}}{P} \sum_{p=1}^{P} w^{(k,p)} e^{j(\varphi^{(k,m,p)} + \psi^{(k,p)})}, 1 \leq m \leq M, \tag{3}$$

where, $\beta_k/P$ is the gain (modeled as a complex random gaussian variable under Rayleigh fading assumption, with a total unitary power over the entire RIS, i.e. with $\mathrm{E}\left[|\beta^{(k)}|^2\right] = 1$) of the path between the 1st antenna element of the BS and the target UE, passing by the $p^{th}$ element of the $k^{th}$ RIS, $w_{k,p}$ is the phase-shift weight of the $p^{th}$ element of the $k^{th}$ RIS, $\varphi_{k,m,p}$ and $\psi_{k,p}$ are phase-shifts defined hereafter:

$$\varphi^{(k,m,p)} = \frac{2\pi}{\lambda}\left(\overrightarrow{a^{(k)}}\cdot\overrightarrow{v^{(m)}} + \overrightarrow{a^{(k)}}\cdot\overrightarrow{c^{(k,p)}}\right), \quad (4)$$

$$\psi^{(k,p)} = \frac{2\pi}{\lambda}\left(\overrightarrow{b^{(k)}}\cdot\overrightarrow{c^{(k,p)}}\right), \quad (5)$$

where, $\overrightarrow{c^{(k,p)}}$ is the vector between the positions of the first and the $p^{th}$ element of the $k^{th}$ RIS, $\overrightarrow{a^{(k)}}$ and $\overrightarrow{b^{(k)}}$ are the unitary vectors indicating the BS-to-$k^{th}$ RIS direction and the $k^{th}$ RIS-to-target UE direction, respectively. Finally, we consider a position Q close to the BS. For such position, we assume a free-space propagation and a spherical wave model. With these assumptions, the channel vector $\mathbf{q}(Q) \in \mathbb{C}^{1\times M}$ between the BS and the UE Q is given by:

$$\mathbf{q}_m(Q) = \frac{\lambda}{4\pi d^{(m)}(Q)}e^{j2\pi\frac{d^{(m)}(Q)}{\lambda}}, \quad m \le M. \quad (6)$$

where, $d^{(m)}(Q)$ is the distance between the $m^{th}$ element of the BS and the UE Q.

*B. RIS-assisted BF procedure, received powers at T and Q*

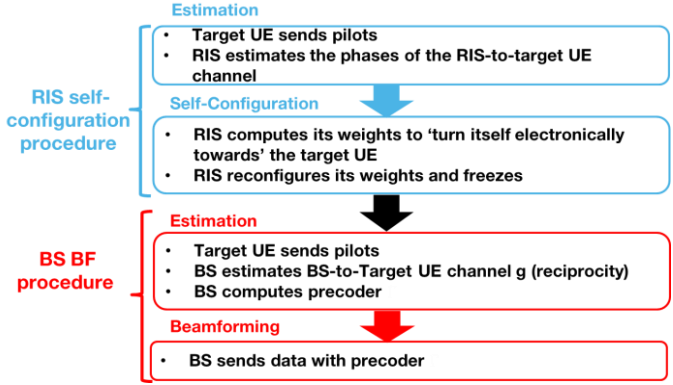

Fig. 3. RIS-assisted Beamforming Procedure

The RIS-assisted BF procedure [11] is illustrated in Fig. 3 and is composed of two phases illustrated in Fig. 3, and detailed hereafter. Note that the BF procedure without RIS assistance simply consists of the second phase only.

During the RIS self-configuration phase, the target UE sends pilots in the uplink, to allow any RIS $k$ to estimate the target UE-to-RIS channel phases $\psi^{(k,p)}$'s. We consider a RIS with continuous phase-shifting capability, as the prototypes used in [17][18]. Hence, each RIS can then computes weights $w_{k,p}$'s to 'turn itself electronically' to the target UE, as follows:

$$w^{(k,m,p)} = e^{-j\psi^{(k,p)}}. \quad (7)$$

Then, each RIS reconfigures its weights, and 'freezes".

During the BS BF phase, the target UE sends pilots in the uplink again, to allow the BS to measure the target UE-to-BS channel. The BS exploits channel reciprocity and deduces the BS-to-Target UE channel $\mathbf{g}$. Finally, the BS computes the unitary beamforming vector $\mathbf{b} \in \mathbb{C}^{M\times 1}$ (with $\|\mathbf{b}\| = 1$) based on $\mathbf{g}$. The BS transmit power is denoted by $\chi \le \chi^{max}$. The received power $\rho$ at the target UE T and the received power $\omega(Q,\mathbf{b},\chi)$ at the UE Q close to the BS are given by:

$$\rho = |\mathbf{gb}|^2\chi, \quad (8)$$

$$\omega(Q,\mathbf{b},\chi) = |\mathbf{q}(Q)\mathbf{b}|^2\chi. \quad (9)$$

Depending on the BF scheme, the expressions of $\mathbf{b}$ and $\chi$ differ. We denote by $\mathbf{b}^{MRT}$, $\mathbf{b}^{red}$, $\mathbf{b}^{trunc}$ and $\mathbf{b}^{boost}$ the precoders for the MRT, the reduced MRT, the Truncated MRT scheme, the Truncated and Boosted MRT schemes, respectively. We denote by $\chi^{MRT}$, $\chi^{red}$, $\chi^{trunc}$ and $\chi^{boost}$ the transmit powers for the MRT, the reduced MRT, the Truncated MRT scheme, the Truncated and Boosted MRT schemes, respectively.

The computation of the precoders $\mathbf{b}^{trunc}$ and $\mathbf{b}^{boost}$ requires the projection of $\mathbf{b}^{MRT}$ onto a Discrete Fourier Transform (DFT) beamforming codebook. Such projection allows for the manipulation of the radiation pattern in some directions (corresponding to some vectors of the codebook), individually, without impacting others. More precisely, we define $\mathbf{F} \in \mathbb{C}^{M\times M}$ as the DFT codebook matrix, where the $m^{th}$ column vector $\mathbf{f}^{(m)} \in \mathbb{C}^{M\times 1}$ is a BF vector pointing towards one distinct direction, among $M$ available directions. With this definition,

$$\mathbf{f}_l^{(m)} = M^{-\frac{1}{2}}e^{j2\pi\frac{(l-1)(m-1)}{M}}, \quad 1 \le l \le M. \quad (10)$$

To get an indication of the main direction of $\mathbf{f}^{(m)}$, we determine the position $Q^{(m)}$ upon the limit circle $\mathcal{C}$, that receives the maximum power, independently from $\chi$:

$$Q^{(m)} = \arg\left\{\max_{Q\subset\mathcal{C}}\{\omega(Q,\mathbf{f}^{(m)},\chi)/\chi\}\right\}. \quad (11)$$

The position $Q^{(m)}$ provides a good indication of the direction of the beam $\mathbf{f}^{(m)}$ of the codebook, as long as $R \gg \lambda$.

We split the limit circle into arcs centered on each $Q^{(m)}$. The separation between two consecutive arcs $S^{(m)}$ and $S^{(m+1)}$ being at mid-distance between $Q^{(m)}$ and $Q^{(m+1)}$.

*C. MRT BF scheme*

For the MRT BF [3], $\mathbf{b}^{MRT}$ and $\chi^{MRT}$ are simply given by:

$$\mathbf{b}^{MRT} = \mathbf{g}^\dagger/\|\mathbf{g}^\dagger\| \text{ and } \chi^{MRT} = \chi^{max}. \quad (12)$$

*D. Reduced MRT BF scheme*

For the reduced MRT BF [11],

$$\mathbf{b}^{red} = \mathbf{b}^{MRT} \text{ and } \chi^{red} = \min\left(\frac{\omega^{thresh}}{\omega^{max}}\chi^{max},\chi^{max}\right), \quad (13)$$

where $\omega^{max}$ is the maximum power received on the limit circle $\mathcal{C}$ and defined as follows: $\omega^{max} = \max_{Q\subset\mathcal{C}}\{\omega(Q,\mathbf{b}^{MRT},\chi^{max})/\chi^{max}\}$. $\omega^{max}$ is determined numerically by the BS based on (9) and (6). Note that, in the case where MRT BF leads the over-exposed area to exceeding the limit circle, then $\omega^{max} > \omega^{thresh}$, and thus $\chi^{red} < \chi^{max}$. A reduced MRT BF 'reduces' its transmit power compared to MRT BF, to get the over-exposed area inside the limit circle.

*E. Truncated MRT BF scheme*

For the Truncated MRT BF scheme, we implement a slightly improved version of [11]. The projection $\boldsymbol{\gamma}^{MRT} \in \mathbb{C}^{M \times 1}$ of $\mathbf{b}^{MRT}$ over $\mathbf{F}$ is first computed: $\boldsymbol{\gamma}^{MRT} = \mathbf{F}^T \times \mathbf{b}^{MRT}$. Such projection aims at truncating the radiation pattern in one direction, individually, with minimum impact on others. Then, the set $\mathcal{M}$ of directions exceeding the limit circle $\mathcal{C}$, is determined as follows:

$$m \in \mathcal{M} \Leftrightarrow \exists Q \subset S^{(m)} | \omega(Q \subset, \mathbf{b}^{MRT}, \chi^{max}) > \omega^{thresh}. \quad (17)$$

The maximum received power over the arc $S^{(m)}$ is given by:

$$\omega^{trunc,m} = \max_{Q \subset S^{(m)}} \{\omega(Q, \mathbf{b}^{MRT}, \chi^{max})\}.$$

A new vector $\boldsymbol{\gamma}^{trunc} \in \mathbb{C}^{M \times 1}$ is computed, to truncate only directions of the radiation pattern that exceed the limit circle $\mathcal{C}$:

$$\boldsymbol{\gamma}_m^{trunc} = \boldsymbol{\gamma}_m \sqrt{\frac{\omega^{thresh}}{\omega^{trunc,m}}} \text{ if } m \in \mathcal{M}, \quad (18)$$

$$\boldsymbol{\gamma}_m^{trunc} = \boldsymbol{\gamma}_m, \text{ otherwise.}$$

Finally, the precoder and the transmit power are derived:

$$\mathbf{b}^{trunc} = (\|\mathbf{F}\boldsymbol{\gamma}^{trunc}\|)^{-1} \mathbf{F}\boldsymbol{\gamma}^{trunc}. \quad (19)$$

$$\chi^{trunc} = \chi^{max} \left(\frac{\|\mathbf{F}\boldsymbol{\gamma}^{trunc}\|}{\|\mathbf{F}\boldsymbol{\gamma}\|}\right)^2. \quad (20)$$

*F. Novel Truncated & Boosted MRT BF scheme*

The novel Truncated and Boosted MRT BF scheme, adds a boosting step to the Truncated BF precoder computation. The set $\mathcal{N}$ of directions (of the radiation pattern) remaining inside the limit circle $\mathcal{C}$, is determined as follows:

$$m \in \mathcal{N} \Leftrightarrow \exists Q \subset S^{(m)} | \omega(Q^{(m)}, \mathbf{b}^{trunc}, \chi^{trunc}) < \omega^{thresh} \quad (21)$$

The maximum received power over the arc $S^{(m)}$ is given by:

$$\omega^{boost,m} = \max_{Q \subset S^{(m)}} \{\omega(Q^{(m)}, \mathbf{b}^{trunc}, \chi^{trunc})\}.$$

A new vector $\boldsymbol{\gamma}^{boost} \in \mathbb{C}^{M \times 1}$ is computed to boost only directions exceeding the limit circle $\mathcal{C}$. More precisely, $\boldsymbol{\gamma}^{boost}$ I initialized to $\boldsymbol{\gamma}^{red}$. Then, it is iteratively updated for each $m \in \mathcal{N}$, as follows, as long as $\chi^{max}\left(\frac{\|\mathbf{F}\boldsymbol{\gamma}^{boost}\|}{\|\mathbf{F}\boldsymbol{\gamma}\|}\right)^2 < \chi^{max}$:

$$\boldsymbol{\gamma}_m^{boost} = \boldsymbol{\gamma}_m \sqrt{\frac{\omega^{thresh}}{\omega^{boost,m}}}. \quad (22)$$

Finally, the precoder and the transmit power are derived:

$$\mathbf{b}^{boost} = (\|\mathbf{F}\boldsymbol{\gamma}^{boost}\|)^{-1} \mathbf{F}\boldsymbol{\gamma}^{boost}, \quad (23)$$

$$\chi^{boost} = \min\left(\chi^{max}\left(\frac{\|\mathbf{F}\boldsymbol{\gamma}^{boost}\|}{\|\mathbf{F}\boldsymbol{\gamma}\|}\right)^2, \chi^{max}\right). \quad (24)$$

## III. VISUALISATION OF THE IMPACT OF BF SCHEMES UPON THE OVER-EXPOSED-AREA

In this Section, we propose to visualize the impact of the studied BF schemes, over the over-exposed area (i.e. the area where $\omega > \omega^{threshold}$), for a given random propagation channel sample. For each BF scheme with a precoder $\mathbf{b}$ and a transmit power $\chi$, we calculate the received power $\omega(Q, \mathbf{b}, \chi)$ at locations Q near the BS, using the mathematical expressions given in Section II. The obtained results correspond to one random channel sample. The simulation parameters are fixed as follows: $M = 64, N = 3, K = 3, P = 16, R = 650$ (this value being the radius normalized by $\lambda$ and $\frac{\omega^{thresh}}{\chi^{max}} = -70\text{dB}$. Throughout this section, all distances and powers are normalized by $\lambda$ and $\chi^{max}$, respectively. We consider two-dimensional (2D) propagation in the x-y plane. The linear array of the base station antenna is at coordinate (0, 0) and is deployed along the x-axis. The BS-to-scatterers and BS-to-RISs directions are plotted in Fig. 4-a) and Fig. 4-b), respectively.

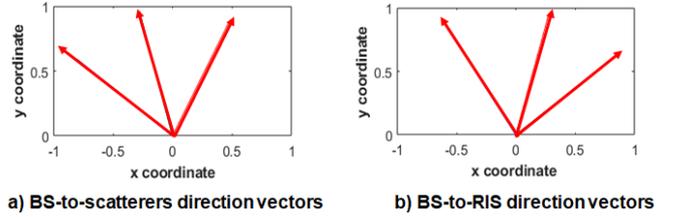

a) BS-to-scatterers direction vectors     b) BS-to-RIS direction vectors

Fig. 4. BS-to-scatterers and BS-to-RIS directions

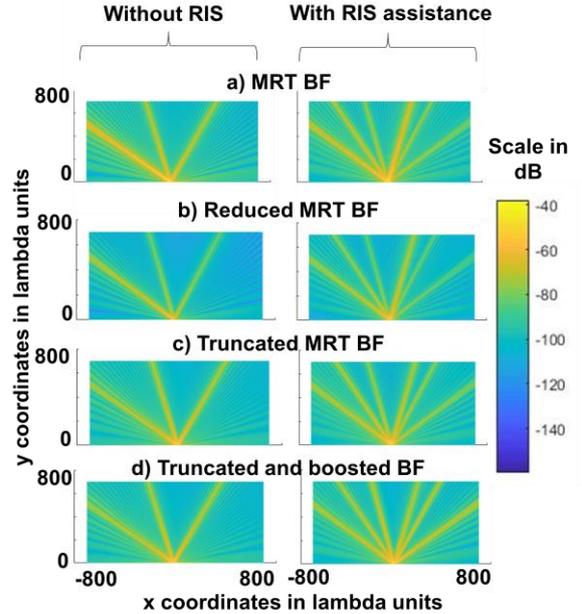

Fig. 5. Received power $\omega$ close to the BS, for all BF schemes

In Fig. 5, to visualize the radiation pattern, we first plot the received power $\omega$ (in dB) at positions near the BS for all the studied. As expected, as they are all derived from MRT BF, all BF schemes radiate towards the scatterers and the RISs, i.e. in the same directions as those depicted in Fig. 4. We can also notice that the Reduced BF scheme severely reduces the radiation pattern in all directions. In contrast, the Truncated BF scheme only reduces the pattern in some directions. Compared to the Truncated BF scheme, the Truncated and Boosted BF scheme boosts some directions. Finally, we can also clearly observe that the RISs spread the radiation pattern in the angular domain regardless of the considered BF scheme.

In Fig. 6, we plot in red the limit circle $\mathcal{C}$ and in yellow the over-exposed area (i.e. the area where $\omega > \omega^{thresh}$). We can observe that for all schemes, the over-exposed area stretches itself towards scatterers and RISs and that the shape of the area reminds the radiation pattern of Fig. 5. We can also clearly visualize that: (i) with MRT, the over-exposed area exceeds the limit circle only in certain directions (hence such scheme would not be deployed); (ii) the Reduced BF scheme reduces the over-exposed area equally in all directions (it is an homothety) to ensure that the entire over-exposed area remains within the limit circle; (iii) the Truncated BF scheme truncates the directions crossing the circle, and manages to bring them perfectly upon the circle; (iv) the Truncated and Boosted BF scheme boosts directions that are inside the circle until they meet the circle.

TABLE I.  NORMALIZED RECEIVED POWER AT TARGET UE AND PERCENTAGE OF POSITIONS (OUTSIDE THE LIMIT CIRCLE AND INSIDE A SQUARE OF $1400\lambda$ BY $1400\lambda$ AROUND THE BS) EXCEEDING THE THRESHOLD, ASSUMING THE BS SERVES THE SAME UE FOR A LONG PERIOD

| RIS | BF Scheme | Received Power at Target UE (dB) | Percentage of positions exceeding the threshold |
|---|---|---|---|
| No | MRT | 23.5 | 3.6 |
| | Reduced MRT | 14.8 | 0 |
| | Truncated | 17.7 | 0 |
| | Truncated & Boosted | | 0 |
| Yes | MRT | 28 | 3.2 |
| | Reduced MRT | 22 | 0 |
| | Truncated | 24.3 | 0 |
| | Truncated & Boosted | 25 | 0 |

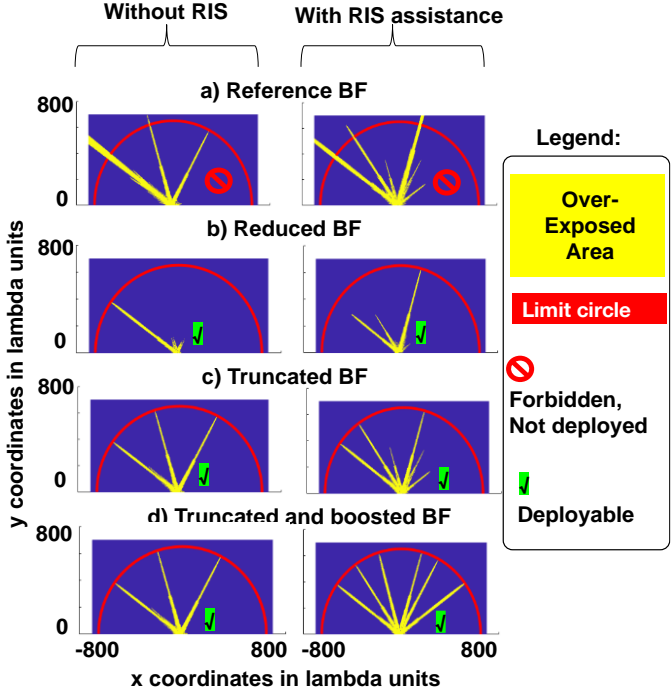

Fig. 6. Over-exposed area (yellow) and limit circle (red) for all BF schemes, assuming the same UE is targeted for a long period.

Table I reports the received power at the target UE and the percentage of positions exceeding the $\omega^{thresh}$ (in the scanned area beyond the limit circle and inside a $1400\lambda$ by $1400\lambda$ square around the BS). Table I shows that Truncated and Boosted MRT BF maximizes the received power, whilst complying with the EMFE constrain, whereas MRT could not be used.

Fig. 6 and Table I also enable to visualize the impact of RIS-assistance. As we can observe in Fig. 6, in this example, when there is no RIS, the over-exposed area of the initial MRT scheme exceeds the circle in the three directions. In this case, truncation is sufficient and additional boosting does not bring any gain in received power, as shown in Table I. In contrast, as we can still observe in Fig. 6, when RISs are introduced, the over-exposed area of the initial MRT scheme contains directions that are inside the circle. In this case, a truncation, is not sufficient, and the Truncated and Boosted BF scheme brings a gain in received power, as shown in Table I.

## IV. STATISTICAL RESULTS

In this section, all schemes are evaluated statistically over random channels. This time, we draw $Nsamples = 1000$ random channel samples. For each sample and for each scheme, we compute (using the model described in Section II) the following metrics: the percentage of positions beyond the limit circle and inside a $1400\lambda \times 1400\lambda$ square around the BS that exceed the threshold; the transmit power $\chi$ and the received power $\omega$ at the target UE. Then, the cumulative density function (CDF) is computed over all samples, for each metric. The simulation parameters are fixed as follows: $M = 64$, $N = 3$, $K = 3$, $P = 16$, $R = 650$ (where $R$ is normalized by $\lambda$) and $\frac{\omega^{threshold}}{\chi^{max}} = -70$dB. Again, all distances and powers are normalized by $\lambda$ and $\chi^{max}$, respectively. The BS linear antenna array is along the x-axis. The directions from the BS to the scatterers and RISs are randomly distributed in $\left[-\frac{\pi}{6}; \frac{\pi}{6}\right]$ with respect to the y-axis. This deployment and propagation scenario is chosen to challenge the BF schemes in terms of EMFE, since it concentrates the scatterers in a narrow angular region.

Fig. 7-a), Fig. 7-b) and Fig. 7-c) present the CDFs of the previously mentioned metrics. The most significant received power on the target UE is obtained with the MRT BF scheme with a violation of the EMFE constraint. Hence, such scheme would not be deployed. On the contrary, this constrain is perfectly met by the Reduced BF scheme, but at the expense of the received power at the target UE, due to a weaker BS transmit power. The Truncated schemes improves the received power (compared to the Reduced BF) whilst exactly meeting the EMFE constrain. As expected, the Truncated and Boosted BF scheme further improve the received power, compared to the Truncated BF scheme. However, this received power gain comes at the expense of a large transmit power increase. Indeed, the BS strongly boosts its power in the direction of weak propagation paths. Also, whereas the exposure constrain is perfectly met with the Reduced and the Truncated schemes, with the Truncated and Boosted BF scheme, less than 0.5% of positions which are over-exposed for less than 50% of samples. Indeed, when the same propagation path impacts two neighbouring arcs, it potentially leads to the reduction of two neighboring beams (and exposure over-reduction of the power) or the boosting of two neighbouring beams (and exposure over-boosting).

As a conclusion, Truncated and Boosted BF maximizes the QoS. However, it consumes more energy. Advantageously, Truncated BF is the best trade-off between QoS and energy-efficiency, and perfectly complies with the EMFE constrain.

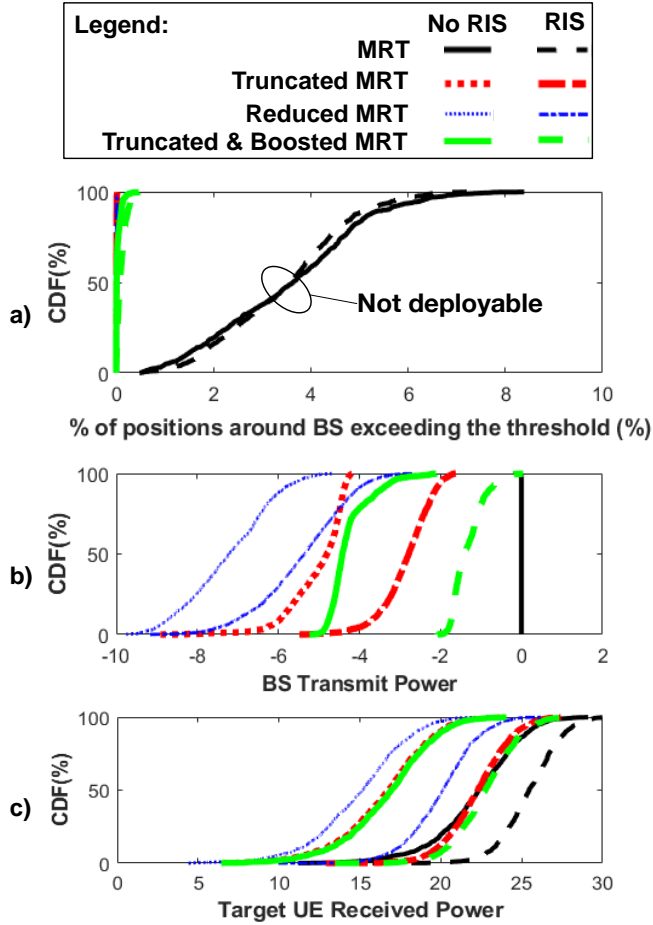

Fig. 7. CDFs of a) positions around BS (inside a 1400 $\lambda$ by 1400 $\lambda$ square around the BS) exceeding the threshold b) BS transmit power and c) target UE received, assuming the same UE is targeted for a long period.

## V. CONCLUSION

In this paper, we propose a novel beamforming scheme called "Truncated and Boosted Maximum Ratio Transmission" beamforming assisted by self-tuning reconfigurable intelligent surfaces. Compared to Maximum Ratio Transmission beamforming, the proposed new system modifies its radiation pattern in such way that the over-exposed area is (i) spread in the angular domain, (ii) truncated in directions exceeding the limit circle (where the received power should not exceed the threshold defined by regulation) and (iii) boosted in directions inside the circle. Our simulations show that the proposed novel scheme maximizes the received power at the target. However, truncation alone is the best trade-off between received power and energy efficiency, under exposure constrain. Future works will take into account more realistic propagation models.

## ACKNOWLEDGMENTS

This work was partially conducted within the framework of the European Union's Horizon 2020 research and innovation project RISE-6G under EC Grant 101017011. We thank our colleagues Yuan-Yuan Huang and Dominique Nussbaum for the good discussions on exposure.

## REFERENCES

[1] A. Morgado, K. M. S. Huq, S. Mumtaz and J. Rodriguez, "A Survey of 5G Technologies: Regulatory, Standardization and Industrial Perspectives," Digital Comm. and Net., no. 4, pp. 87-97, 2018.

[2] F. Rusek et al., "Scaling up MIMO: Opportunities and Challenges with Very Large Arrays," IEEE Sig. Proc. Mag., vol. 30, no. 1, pp. 40-60, Jan. 2013.

[3] T. K. Y. Lo, "Maximum Ratio Transmission," IEEE Trans. on Comm., vol. 47, no. 10, pp. 1458-1461, Oct. 1999.

[4] F. W. Vook, A. Ghosh and T. A. Thomas "MIMO and Beamforming Solutions for 5G Technology," IEEE MTT-S IMS, pp. 1-4, June 2014.

[5] B. Thors, A. Furuskär, D. Colombi and C. Tornevik "Time-Averaged Realistic Maximum Power Levels for the Assessment of Radio Frequency Exposure for 5G Radio Base Stations Using Massive MIMO,'" IEEE Access, vol. 5, pp. 19711-19719, Sept. 2017.

[6] E. Degirmenci, B. Thors and C. Tornevik, "Assessment of Compliance With RF EMF Exposure Limits: Approximate Methods for Radio Base Station Products Utilizing Array Antennas With Beam-Forming Capabilities," IEEE Transactions on Electromagnetic Compatibility, vol. 58, no. 4, pp. 1110-1117, August 2016.

[7] B. Xu, K. Zhao, Z. Ying, D. Sjöberg, W. He and S. He, "Analysis of Impacts of Expected RF EMF Exposure Restrictions on Peak EIRP of 5G User Equipment at 28 GHz and 39 GHz Bands," IEEE Access, vol. 7, pp. 20996-21005, February 2019.

[8] S. Persia, C. Carciofi, S. D'Elia and R. Suman, "EMF Evaluations for Future Networks based on Massive MIMO," Int'l Symp. on Personal, Indoor and Mobile Radio Comm. (PIMRC), pp. 1197-1202, Sept. 2018.

[9] D. Colombi et al., "Assessment of Actual Maximum RF EMF Exposure from Radio Base Stations with Massive MIMO Antennas," PhotonIcs \& Electromagnetics Research Symp. - Spring, pp. 570-577, June 2019.

[10] L. Chiaraviglio, A. Elzanaty and M. -S. Alouini "Health Risks Associated With 5G Exposure: A View From the Communications Engineering Perspective, " in IEEE Open Journal of the Communications Society, vol. 2, pp. 2131-2179, 2021.

[11] N. Awarkeh, D.-T. Phan-Huy, R. Visoz, "Electro-Magnetic Field (EMF) aware beamforming assisted by Reconfigurable Intelligent Surfaces," IEEE 22nd International Workshop on Signal Processing Advances in Wireless Communications (SPAWC), pp. 541-545, Sept 2021.

[12] M. D. Renzo, et al. "Smart Radio Environments Empowered by Reconfigurable AI Meta-Surfaces: an Idea whose Time has Come," EURASIP J. on Wireless Comm. and Net., May 2019.

[13] E. Basar, M. D. Renzo, J. De Rosny, M. Debbah, M.S. Alouini and R. Zhang, "Wireless Communications through Reconfigurable Intelligent Surfaces," IEEE Access, vol. 7, pp. 116753-116773, 2019.

[14] E. C. Strinati et al., "Wireless Environment as a Service Enabled by Reconfigurable Intelligent Surfaces: The RISE-6G Perspective," 2021 Joint EuCNC/6G Summit, 2021, pp. 562-567.

[15] H. Ibraiwish, A. Elzanaty, Y. H. Al-Badarneh and M. -S. Alouini "EMF-Aware Cellular Networks in RIS-Assisted Environments," in IEEE Communications Letters, vol. 26, no. 1, pp. 123-127, Jan. 2022.

[16] A. Zappone, M. Di Renzo "Energy Efficiency Optimization of Reconfigurable Intelligent Surfaces with Electromagnetic Field Exposure Constraints," submitted to Signal Proc. Letters, Jan. 2022.

[17] R. Fara et al., "Reconfigurable Intelligent Surface-Assisted Ambient Backscatter Communications – Experimental Assessment," 2021 IEEE ICC Workshops, 2021, pp. 1-7.

[18] R. Fara et al., "A Prototype of Reconfigurable Intelligent Surface with Continuous Control of the Reflection Phase," in IEEE Wireless Communications, vol. 29, no. 1, pp. 70-77, February 2022.